\documentclass[twocolumn,showpacs,prl,amsmath,amssymb,superscriptaddress]{revtex4}
\usepackage{graphicx}
\usepackage{dcolumn}
\usepackage{bm}
\usepackage[latin1]{inputenc}
\usepackage[mathscr]{eucal}
\usepackage{epsfig}
\usepackage{rotating}
\usepackage{multirow}

\begin{document}

\title{Electronic Refrigeration at the Quantum Limit}

\author{Andrey V. Timofeev} \affiliation{Low Temperature
Laboratory, Helsinki University of Technology, P.O. Box 3500,
02015 TKK, Finland} \affiliation{Laboratory of Superconductivity,
Institute of Solid State Physics, Chernogolovka, 142432 Russia}

\author{Meri Helle} \affiliation{Low Temperature
Laboratory, Helsinki University of Technology, P.O. Box 3500,
02015 TKK, Finland}

\author{Matthias Meschke} \affiliation{Low Temperature
Laboratory, Helsinki University of Technology, P.O. Box 3500,
02015 TKK, Finland}

\author{Mikko M\"ott\"onen} \affiliation{Low Temperature
Laboratory, Helsinki University of Technology, P.O. Box 3500,
02015 TKK, Finland} \affiliation{Department of Applied Physics,
Helsinki University of Technology, P.O. Box 5100, 02015 TKK,
Finland} \affiliation{Australian Research Council Centre of
Excellence for Quantum Computer Technology, The University of New
South Wales, Sydney 2052, Australia}

\author{Jukka P. Pekola} \affiliation{Low Temperature
Laboratory, Helsinki University of Technology, P.O. Box 3500,
02015 TKK, Finland}

\begin{abstract}
We demonstrate quantum limited electronic refrigeration of a
metallic island in a low temperature micro-circuit. We show that
matching the impedance of the circuit enables refrigeration at a
distance, of about 50 $\mu$m in our case, through superconducting
leads with a cooling power determined by the quantum of thermal
conductance. In a reference sample with a mismatched circuit this
effect is absent. Our results are consistent with the concept of
electromagnetic heat transport. We observe and analyze the
crossover between electromagnetic and quasiparticle heat flux in a
superconductor.
\end{abstract}

\maketitle

The fundamental limit of heat transport via a single channel is
governed by the quantum of thermal conductance \cite{pendry83}.
This phenomenon was verified experimentally for phonons
\cite{schwab00,yung02}, electrons \cite{chiatti06}, and photons
\cite{meschke06}. In the experiment by Meschke {\it et al.}
\cite{meschke06}, the contribution of heat conductance by photons
was relatively weak due to impedance mismatch in the employed
electrical circuit and due to strong electron-phonon coupling.
Here, we demonstrate the importance of matching the circuit to
reach the full quantum of heat conductance. Our experiment allows
for direct observation of heat transport at the limit of one
quantum. We also observe and analyze how two parallel heat
conduction mechanisms in a superconductor
--- by quasiparticles and by thermal photons --- dominate in
different temperature regimes.

As discovered in 1928 by Johnson and Nyquist
\cite{johnson,nyquist}, a resistor $R$ in an electric circuit at
temperature $T$ produces thermal voltage noise with power spectrum
given by $4k_BTR$ per unit frequency bandwidth. Therefore, two
resistors $R_1$ and $R_2$ at different temperatures $T_1$ and
$T_2$ exchange energy in a circuit with a net heat flux between
them: heat flows from hot to cold according to the second law of
thermodynamics. The heat flux discussed here is electromagnetic in
nature \cite{phonons}, and it can be written as \cite{schmidt04}:
\begin{equation} \label{eq1}
P_{\nu}=\int_{0}^{\infty}\frac{d\omega}{2\pi}
\frac{4R_1R_2\hbar\omega}{|Z_t(\omega)|^2}\left(\frac{1}{e^{\hbar
\omega/k_BT_2}-1}-\frac{1}{e^{\hbar \omega/k_BT_1}-1}\right).
\end{equation}
Here, $Z_t$($\omega)$ is the frequency $\omega/2\pi$ dependent
total series impedance of the circuit. Whether the heat exchange
in a circuit is classical (as in \cite{johnson,nyquist}) or
quantum limited depends fundamentally on temperature $T$ and on
the (linear) size of the circuit $\ell$, or, more precisely,
whether the electromagnetic noise, mediating the heat between the
two resistors, is cut-off at the characteristic frequencies of the
circuit $\omega_c=(RC)^{-1}$ or $\omega_c=R/L$, or at the thermal
frequency $\omega_T=k_BT/\hbar$. Rough estimates of unavoidable
stray capacitances $C$ and series inductances $L$ are given by
$C\sim \epsilon \ell$ and $L\sim \mu \ell$, where $\epsilon$ is
the permittivity and $\mu$ the permeability of the medium. For a
macroscopic room temperature $T=300$ K circuit of $\ell\sim 1$ mm
size, the noise is cut-off at the circuit frequency,
$\omega_c/\omega_T \sim 10^{-2}\ll 1$ for the resistance $R=100$
$\Omega$, which is of the same order as that in our experiment. In
this case the noise and the heat flux are classical, originating
from the equipartition law, where each degree of freedom carries
an energy $k_BT$ on the average \cite{nyquist}. Here, the
magnitude of the heat flux does not follow any universal
dependence, but it is determined by the detailed circuit topology
and impedances. For a low temperature micro-circuit as in our
experiment, with $T=100$ mK and $\ell \sim 100$ $\mu$m, we are in
the quantum limit: $\omega_c/\omega_T \sim 10^{2}\gg 1$. In this
case, the heat flux is governed by the equilibrium thermal
distribution of electromagnetic radiation of the resistor and is
limited by the universal quantum of thermal conductance $G_Q
\equiv \pi k_B^2T/6\hbar$. This electromagnetic heat conduction
mechanism dominates in electronic nanostructures
\cite{schmidt04,ojanen07,ojanen08,segal08} over electron-phonon
and normal electronic heat conduction as the temperature
approaches zero.

\begin{figure}
\begin{center}
\includegraphics[width=0.97\linewidth, keepaspectratio]{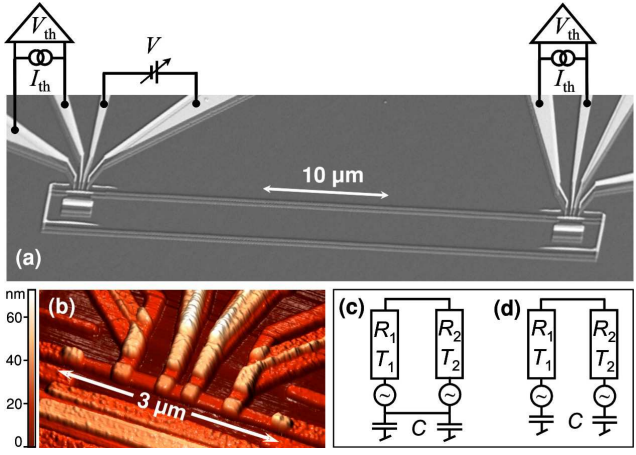}
\end{center}
\caption{(a) Electron micrograph of sample A. Two gold-palladium
Au$_{0.75}$Pd$_{0.25}$ islands at a 50 $\mu$m distance are
connected with aluminium superconducting lines into a loop to
match the impedance between them and enable remote refrigeration.
(b) Colored atomic force microscopy image of the island, connected
at both ends directly to superconducting lines by metal-to-metal
contacts. The four NIS junctions, contacting each island in the
middle part, are used to perturb and to measure the island
temperature. (c) Equivalent electrical circuit of the matched
(sample A) and (d) the mismatched (sample B) structure.}
\label{fig1}
\end{figure}
To observe quantum limited refrigeration and to demonstrate the
significance of impedance matching, we have devised a circuit
shown in Fig. 1a. Two gold-palladium (Au$_{0.75}$Pd$_{0.25}$)
normal metal islands on an oxidized silicon substrate at a
distance $\ell=50$ $\mu$m are connected into a loop by aluminium
superconducting lines. The aluminium and gold-palladium metals are
in a direct contact to each other, without a tunnel barrier,
whereby the contact resistance is small, $< 1$ $\Omega$. Each
island is 3 $\mu$m long, 0.2 $\mu$m wide and 20 nm thick (see Fig.
1b) and the measured resistance of each of them is $R\simeq 230$
$\Omega$. Each island is also connected to four external aluminium
superconducting leads through aluminium oxide tunnel barriers,
which form four normal-metal - insulator - superconductor (NIS)
tunnel junctions, with the area of $150\times150$ nm$^2$ and
measured normal state resistance $R_T\simeq 19$ ${\rm k}\Omega$
each. Different pairs of these junctions are used to perturb and
to measure the electronic temperature of the islands, as detailed
below. We have also fabricated and measured a similar reference
sample, in which the two islands were connected only by a single
superconducting aluminium line and were not enclosed into a loop.
The two sample geometries in the experiment represent impedance
matched (with loop geometry) and mismatched electrical circuits,
which are schematically depicted in Figs. 1c and 1d, and which we
denote sample A and sample B, respectively.

To show that impedance matching between the two islands is indeed
vital for the observation of quantum limited refrigeration, we
compare the rates $P_\nu$ in the matched and mismatched cases. For
the matched sample A with $R_1=R_2=R,~ Z_t(\omega)=2R$, we obtain
from Eq. \eqref{eq1} the universal quantum heat flux: $P_\nu ^{\rm
A} = \frac{\pi k_B^2}{12\hbar}(T_2^2 - T_1^2)$, which is $G_Q
(T_2-T_1)$ for a small temperature difference and presents the
maximum heat flux possible for transmission through this
electromagnetic channel. For the mismatched circuit with
$R_1=R_2=R$, closed by the shunt capacitance $C/2\sim 10$ fF which
is determined mainly by the NIS junctions, we obtain $P_{\nu}^{\rm
B}\simeq\frac{\pi^3 k_B^2}{30\hbar}(T_2^2 -
T_1^2)(k_BTRC/\hbar)^2$ for $T\approx T_1\approx T_2$ at high
circuit cut-off frequencies $\omega_c\gg\omega_T$. We find that
$P_{\nu}^{\rm B}/P_{\nu}^{\rm
A}\simeq\frac{2\pi^2}{5}(\omega_T/\omega_c)^2$; the
electromagnetic power flow in the matched circuit is expected to
be about $10^2 - 10^3$ times stronger as compared to that in the
mismatched case at temperatures $0.3 - 0.1$ K relevant for the
experiment.

The two samples were fabricated with the standard methods of
electron beam lithography and shadow evaporation, and measured in
a $^3$He-$^4$He dilution refrigerator as follows. One pair of NIS junctions, the
SINIS-refrigerator, connected to island 1, is DC-biased with
voltage $V$ to cool down the island by removing hot electrons from
it into the superconducting leads through the tunnel barrier at
voltages $eV<2\Delta$ \cite{nahum94,leivo96,clark05,giazotto06}. Here $\Delta\simeq 200$
$\mu$eV is the superconducting energy gap of aluminium \cite{gap}.
At higher bias voltages the island is heated up. To probe the
island temperature $T_1$, the other pair of NIS junctions on the
island is used as a thermometer by applying a small DC current
$I_{\rm th}$ through it, and by measuring the corresponding
temperature dependent voltage $V_{\rm th}$. Another
similar SINIS-thermometer probes the temperature $T_2$ of the
second island. When the applied voltage $V$ through the
SINIS-refrigerator is zero, the measured voltage $V_{\rm th}(V=0)$
provides the thermometer calibration against the bath temperature
$T_0$ when the cryostat temperature is varied in the range $50 -
500$ mK. The electronic temperature of the islands is then
obtained from the fit of the dependence of $T_0$ on $V_{\rm
th}(V=0)$. At $V=0$ the electronic temperature coincides with the
bath temperature down to $T_0=120$ mK. The thermometers have
individual floating DC bias sources and do not cause excessive
heating or cooling of the islands due to the low bias current,
$I_{\rm th}\simeq 0.001\Delta/eR_T$, used.
\begin{figure}
\begin{center}
\includegraphics[width=0.85\linewidth,keepaspectratio]{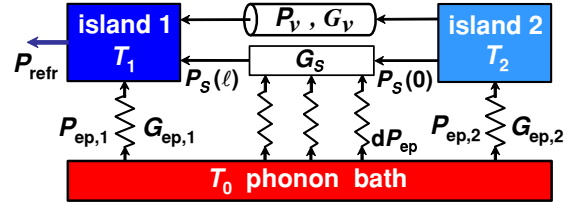}
\end{center}
\caption{The thermal model illustrates electromagnetic ($P_\nu,
G_\nu$) and quasiparticle ($P_s, G_s$) heat conduction through the
superconducting line between the two islands. The islands are
thermally coupled to the phonon bath with the heat fluxes $P_{\rm
ep,1}$ and $P_{\rm ep,2}$. The weak electron-phonon coupling of
the superconducting line to the thermal bath is denoted by ${\rm
d}P_{\rm ep}$. The arrows show the direction of the heat flow for
temperatures $T_1<T_2<T_0$.}\label{fig2}
\end{figure}

The thermal model that accounts for our set-up and observations is
shown in Fig. 2. The resistors exchange energy at power $P_\nu$
through the electromagnetic channel, and, in parallel, at power
$P_s$, due to quasiparticle heat conduction through the
superconducting line. The latter contribution is significant at
higher temperatures but diminishes exponentially towards low
temperatures, $k_BT_0 \ll \Delta$ \cite{bardeen59,timofeev}. We
describe quantitatively quasiparticle heat flux with the heat
diffusion equation
\begin{equation} \label{eq2}
\frac{d}{dx}\left(-\kappa_s \frac {dT}{dx}\right)=\alpha(T_0)
\Sigma_{\rm Al}[T_0^5-T^5(x)],
\end{equation}
assuming that the superconducting line has the temperature profile
$T(x)$ with boundary conditions $T(0)=T_2$ and $T(\ell)=T_1$,
where $x$ is the coordinate along the line ($x=0$ corresponds to
the contact to island 2, $x=\ell$ to that to island 1). Factor
$\alpha(T_0)$ determines the suppression of electron-phonon
coupling in the superconducting line with respect to that in the
normal metal state \cite{timofeev}, and $\Sigma_{\rm
Al}\simeq0.3\cdot10^9$ WK$^{-5}$m$^{-3}$ is a material constant
for aluminium \cite{giazotto06}. The heat flux
$P_s(0)=-\kappa_sAT'(0)$ to island 1, and the heat flux
$P_s(L)=-\kappa_sAT'(\ell)$ from island 2 are determined through
temperature gradients $T'(0)$ and $T'(\ell)$ at the ends of the
superconducting line with cross-sectional area $A=200\times 25$
nm$^2$. Here, $\kappa_s=\gamma(T_0)\kappa_n$ is the heat
conductivity of the superconducting line \cite{bardeen59},
suppressed by a factor
$\gamma(T_0)=\frac{3}{2\pi^2}\int_{\Delta(T_0)/k_BT_0}^{\infty}\frac{t^2dt}{\cosh^2(t/2)}$
with respect to heat conductivity $\kappa_n=\ell
L_0T(x)/(R_{\ell}A)$ in the normal metal state determined by the
Wiedemann-Franz law. Here, $L_0\simeq 2.4\cdot10^{-8}$
W$\Omega$K$^{-2}$ is the Lorenz number, and $R_{\ell}$ is the
normal state resistance of the aluminium line; $R_{\ell} \simeq$
138 $\Omega$ for sample A and $R_{\ell} \simeq$ 188 $\Omega$ for
sample B. The electrons in each resistor of volume $\Omega_{i}$
exchange energy with the substrate, i.e., with the thermal bath at
temperature $T_0$ via electron-phonon coupling at the rate
$P_{{\rm ep},i}=\Sigma_{\rm AuPd}\Omega_i(T_0^5-T_{i}^5)$
\cite{roukes85,wellstood94,volume}, where $\Sigma_{\rm AuPd}\simeq
(2-4)\cdot10^9$ WK$^{-5}$m$^{-3}$ is obtained from the
measurements. Island 1 can be SINIS-refrigerated (or heated) with
the corresponding power $P_{\rm refr}$. We neglect phonon heat
transport based on experimental results discussed below. The
steady-state of the system is then described by the energy balance
equations
\begin{eqnarray}\label{eq3}
P_{\rm refr}-P_{\nu}-P_s(\ell)-P_{{\rm ep},1}=0 \nonumber \\
P_{\nu}+P_s(0)-P_{{\rm ep},2}=0.
\end{eqnarray}
For the quantitative analysis of the remote refrigeration effect,
we solve numerically Eq. \eqref{eq2} together with Eqs.
\eqref{eq3} to obtain the relative temperature change of island 2
with respect to that of island 1, $\Delta T_2/\Delta T_1 \equiv
(T_2-T_0)/(T_1-T_0)$. For small temperature differences,
neglecting the electron-phonon coupling in the superconductor, we
can linearize the different contributions in Eqs. \eqref{eq3} and
obtain a particularly simple expression for $\Delta T_2/\Delta
T_1$:
\begin{equation}\label{eq4}
\frac{\Delta T_2}{\Delta T_1} = \frac{G_\nu+G_s}{G_\nu+G_s+G_{{\rm
ep},2}}.
\end{equation}
Here, the photon coupling $G_\nu$ is expected to be equal to $G_Q$
for the matched sample and for the mismatched sample it is
suppressed by a large factor as discussed above. The
electron-phonon conductance is given by $G_{{\rm
ep},2}=5\Sigma_{\rm AuPd}\Omega_2 T_0^4$, and $G_s$ denotes the
ordinary heat conductance by quasiparticles in the superconducting
line.
\begin{figure}
\begin{center}
\includegraphics[width=\linewidth,keepaspectratio]{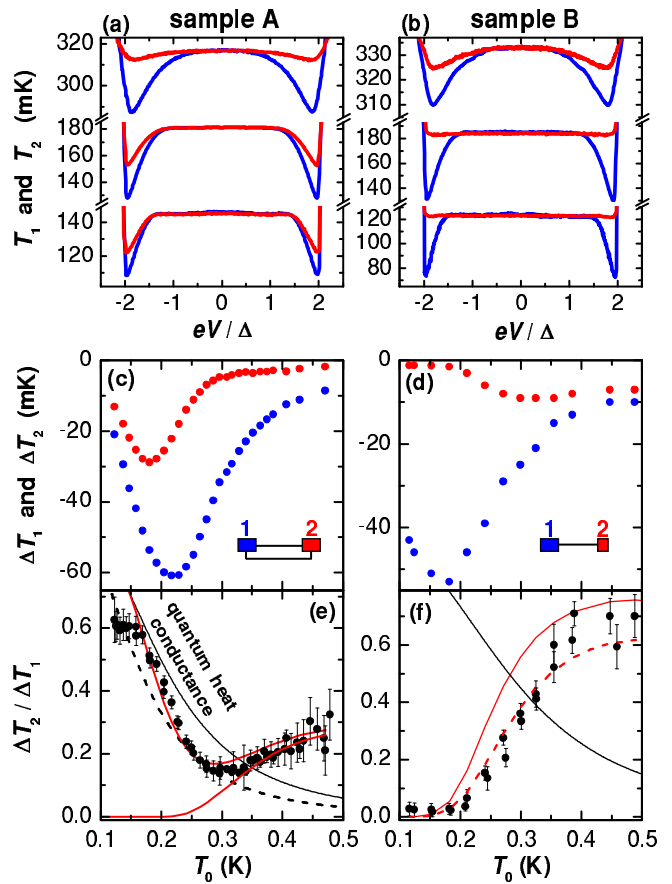}
\end{center}
\caption{Measured data of sample A (a, c, e) and sample B (b, d,
f) and calculated results of the thermal model. (a), (b):
Measured island temperatures $T_1$ (blue line) and $T_2$ (red
line) vs. bias voltage $V$ at a few bath temperatures $T_0$. (c), (d):
Absolute temperature changes $\Delta T_1$ (blue dots) and $\Delta
T_2$ (red dots) measured at bath temperatures 120 - 500
mK. (e), (f): Relative temperature changes obtained at 120
- 500 mK are shown by the black dots. The error bars show the
standard deviation arising from the temperature calibration.
(e), (f): The black lines are obtained from the linearized thermal
model. The red lines are the results of the numerical thermal
model. The solid lines are calculated with $\Sigma_{\rm
AuPd}=2\cdot 10^9$ WK$^{-5}$m$^{-3}$ and the dashed lines with
$\Sigma_{\rm AuPd}=4\cdot 10^9$ WK$^{-5}$m$^{-3}$.}\label{fig3}
\end{figure}

Upon sweeping the voltage $V$ across the SINIS-refrigerator, both
thermometers show cooling at $eV \lesssim 2\Delta$, indicating
electronic refrigeration of both islands, see Figs. 3a and 3b. As
expected, the SINIS refrigerating effect is maximal around the
optimal bias voltage $eV \simeq 2\Delta$. The absolute temperature
drops $\Delta T_1$ and $\Delta T_2$ of the two islands for the two
samples are shown in Figs. 3c and 3d. In both cases, the direct
SINIS refrigeration of island 1 is similar. In contrast, the
refrigeration of the remote island is drastically different in the
two samples. In sample A with the matched circuit, island 2
temperature tends to follow the temperature of the main island at
low temperatures. In sample B with the mismatched circuit, the
refrigeration of the remote island is suppressed at low
temperatures. The corresponding relative temperature drops $\Delta
T_2/\Delta T_1$ against $T_0$ at the optimum cooling bias voltage
are plotted in Figs. 3e and 3f. For the matched sample $\Delta
T_2/\Delta T_1$ has a minimum as a function of $T_0$ at about 300
mK; below this temperature it increases rapidly, as can be
expected based on strong electromagnetic coupling. In the
mismatched sample, $\Delta T_2/\Delta T_1$ vanishes towards lower
temperatures, due to weak photonic coupling.

For the matched sample, the rise of $\Delta T_2/\Delta T_1$ at low
temperatures in Fig. 3e is in agreement with the simple linearized
thermal model: the data below 300 mK lie between the solid and
dashed black lines, obtained from Eq. \eqref{eq4} assuming full
quantum conductance $G_\nu=G_Q$ and vanishing quasiparticle
conductance $G_s=0$. Alternatively, the dashed black curve can be
obtained from Eq. \eqref{eq4} with $G_\nu=0.5G_Q$ and $\Sigma_{\rm
AuPd}=2\cdot10^9$ WK$^{-5}$m$^{-3}$, since only the photonic
$G_{\nu}$ and the electron-phonon $G_{\rm ep,2}$ coupling
contribute to the relative temperature drop $\Delta T_2/\Delta
T_1$ when $G_s=0$. At very low temperatures, the thermometer
signal saturates. Based on these results we conclude that the
matching in this sample is close to ideal and the refrigeration is
limited by the quantum of thermal conductance. This effect is
absent in the mismatched sample: for reference we show the black
line in Fig. 3f with $G_\nu=G_Q$ and $G_s=0$.

The quantitative behaviour of $\Delta T_2/\Delta T_1$ at high
temperatures $T_0\gtrsim 300$ mK is not universal and depends on
sample parameters. In this temperature regime, the remote island
of sample B is refrigerated more than that in sample A. This is
because of stronger thermalization of sample A with larger island
2, and since there are, due to the deposition technique, extra
normal (AuPd) shadows covering the vertical parts of the aluminium
looped line (see Fig. 1a). In sample B, the normal shadow is not
in contact with the superconducting line, which further enhances
the quasiparticle mediated refrigeration. For Sample A, the data
over the full temperature range are accounted for by the upper red
line in Fig. 3e, obtained from the numerical analysis with
$P_{\nu}=P^{\rm A}_{\nu}$ and $\Sigma_{\rm AuPd}=2\cdot10^9$
WK$^{-5}$m$^{-3}$. To fit the data for sample A in the diffusion
regime at temperatures above 300 mK, we added a fitting parameter
$\alpha_N=0.6$ to the factor $\alpha$:
$\alpha\rightarrow\alpha+\alpha_N$. The parameter $\alpha_N$
describes stronger thermalization of the superconducting line. The
lower red line in Fig. 3e, calculated with $\alpha_N=0.6$ and with
no photonic heat exchange $P_{\nu}=0$, shows quasiparticle
contribution for comparison. For sample B, the numerically
obtained red curves of quasiparticle conduction are shown in Fig.
3f. The dashed red curve shows good agreement with the data. The
uncertainty in the quantitative comparison between the model and
data arises from only approximately known parameters of
electron-phonon coupling for gold-palladium and aluminium thin
films.

In both samples, at voltages $eV>2\Delta$, the probe islands are
strongly heated due to hot quasiparticle injection in this regime.
An additional thermometer, located near island 2, but not
connected to it, was monitoring phonon temperature on
the substrate. It showed negligibly weak temperature response as
compared to the thermometers of islands 1 and 2. This supports our
thermal model, which assumes that phonons provide a good thermal
bath and that the observed heat exchange between the resistors
occurs due to quasiparticles and electromagnetic coupling. This is
a natural conclusion due to the very weak electron-phonon coupling
at low temperatures.

In conclusion, we have demonstrated quantum limited refrigeration:
the low temperature data show quantitative agreement with the
thermal model assuming heat conduction determined by the quantum
of thermal conductance. Furthermore, our observations and model
account for residual heat conduction in a superconductor by
quasiparticles. We suggest that even galvanically decoupled
resistors can be refrigerated by the mechanism discussed. This
could be an option for noise suppression purposes in sensitive
quantum devices, e.g., by refrigerating shunt resistors
\cite{wellstood94} in SQUIDs (superconducting quantum interference
devices).

We thank N. Chekurov for technical assistance in AFM imaging. This
work is supported by the Academy of Finland, National Graduate
School for Materials Physics, and the NanoSciERA project
"NanoFridge" of the EU.


\begin{thebibliography}{99}
\bibitem{phonons} There are generally other contributions of heat flux,
which are often dominant, like that between the resistor and the
phonon bath, i.e., the usual conduction to the surroundings, and
electronic heat conduction along the wiring.
\bibitem{gap} The measured value of $\Delta$ is $200\pm 5$ $\mu$eV for sample A
and $210\pm 5$ $\mu$eV for sample B.
\bibitem{volume} The volume $\Omega_2$ of island 2 for sample A is $1.2\cdot10^{-20}$ m$^3$
and for sample B it is $4.3\cdot10^{-21}$ m$^3$.
\end{thebibliography}
\end{document}